# Dynamic inference of user context through social tag embedding for music recommendation

Dynamic inference of user context for music recommendation

Diego Sánchez-Moreno

University of Salamanca, Department of Computer Science and Automation, sanchez91@gmail.com

Álvaro Lozano Murciego

University of Salamanca, Department of Computer Science and Automation, loza@usal.es

Vivian F. López Batista

University of Salamanca, Department of Computer Science and Automation, vivian@usal.es

María Dolores Muñoz Vicente

University of Salamanca, Department of Computer Science and Automation, mariado@usal.es

María N. Moreno-García*

University of Salamanca, Department of Computer Science and Automation, mmg@usal.es

Music listening preferences at a given time depend on a wide range of contextual factors, such as user emotional state, location and activity at listening time, the day of the week, the time of the day, etc. It is therefore of great importance to take them into account when recommending music. However, it is very difficult to develop context-aware recommender systems that consider these factors, both because of the difficulty of detecting some of them, such as emotional state, and because of the drawbacks derived from the inclusion of many factors, such as sparsity problems in contextual pre-filtering. This work involves the proposal of a method for the detection of the user contextual state when listening to music based on the social tags of music items. The intrinsic characteristics of social tagging that allow for the description of items in multiple dimensions can be exploited to capture many contextual dimensions in the user listening sessions. The embeddings of the tags of the first items played in each session are used to represent the context of that session. Recommendations are then generated based on both user preferences and the similarity of the items computed from tag embeddings. Social tags have been used extensively in many recommender systems, however, to our knowledge, they have been hardly used to dynamically infer contextual states.

## 1 INTRODUCTION

Music is one of the domains in which context has an important influence on user preferences. User activity, emotional state, time, location, weather, etc., are contextual factors that condition the choice of songs or artists to play, and therefore should also be taken into account when offering recommendations. In a user session, when listening to music from a streaming service, for instance, many contextual factors can come into play, some of which may be difficult to detect.

Contextual states can be obtained in several ways: explicitly from the user, implicitly from sensors and devices such as those used for listening to music, or by inference from the user interaction with the system [1]. Moreover, the types of contexts can be very varied, with two general categories [2]: environment-related and user-related. The factors that are relevant in the area of music come from both the first type (location, time, weather…) and the second (user activity, emotional state…). In addition to the difficulty of collecting these factors, considering all of them individually in the recommendation methods would lead to problems such as sparsity or a decrease in the reliability of the models as the number of variables involved increases.

In order to overcome the above problems, we propose to use social tags given to items by users to infer the contextual state representative of a conjunction of unknown factors. We take advantage of this type of tags that make up folksonomies and have the property of characterizing items according to multiple dimensions as opposed to the single-dimension classifications of classical taxonomies. In the music domain, social tags that users assign to items represent attributes and characteristics of the music as well as the context for which it is appropriate. Thus, tags corresponding to different dimensions or contextual factors such as user activity, place or mood, among others, are associated with music items. However, the use of tags can cause noise problems due to the absence of vocabulary restrictions in tagging the items. This drawback has been addressed in different ways in the literature [3], [4], with the application of word embedding techniques to tags (tag embedding) being among the most successful [5]. The fact of not processing tags in isolation but taking into account their context makes it possible to reduce noise-causing inconsistencies and redundancies.

In this work, the tags associated with the music played by the user in each session are processed by mean of word embedding techniques and used to perform contextual post-filtering of the recommendations provided by collaborative filtering (CF) methods. In this way, each tag is located at a point in the space given by its tag embedding vector. Using dimensionality reduction techniques, it is possible to visualize the tags in the two-dimensional space, noting that similar tags in terms of music type and context are close to each other. We consider the different regions of space to be representative of different contextual states of the users involving several context dimensions and the type of music they listen to in those states. Post-filtering is based precisely on finding the items whose tags are closest to those of the items being played in the user's session. This allows us to adapt the recommendations to the implicit context of each session, avoiding sparsity problems caused by contextual prefiltering methods.

Social tags have been used in some works to characterize items [6] and users [7] and thus improve the recommendations provided by CF methods. Even tag embedding is being used more recently in these systems [7, 8]. However, we have found few works in the literature in which tags are used to infer context in music recommendation [3, 4], and none in which tag embedding is used for context inference.

The main contribution of this work is precisely the proposal of a method to infer the contextual state of the user when listening to music and to generate recommendations adapted to that context. We focused on the field of music because it is an application domain that is quite different from others due to the fact that the way music is consumed is very different from other products [1].

The rest of the paper is organized as follows. First, a brief section on related work is included. The next section is devoted to describing the proposed method for inferring the context in each user session and making context-aware recommendations. Section 4 describes the experimental study and presents the results. Finally, conclusions are given in section 5.

## 2 RELATED WORK

Context-aware recommender systems have been the focus of extensive research work for many years. In the field of music, these systems acquire great importance since the context has a great influence on the music listened to by the user, while their complexity is greater than those of other domains because it is difficult to identify all the influencing factors [9]. Proposals for music recommendations have been developed in which different contextual factors are considered [1]. Some of the most influential [10] are user activity [11, 12], emotional state [13, 14], location [15] and time [16], although recently, great importance is also being given to the user social context [17].

Social relations, tags, tweets, reviews, and other information from social networks have been used in music recommender systems for many purposes. User implicit feedback, as well as user and item characterization are sometimes got from social tagging. Tags and other social information are combined in [18] for detecting communities in social networks and creating sub-clustering from tags of artists, which are the basis of the proposed recommender system. Some content-based or hybrid approaches take advantage of social tags to find similarities between items, as in [19], where the similarity between songs inferred from tags is used to make recommendations. Some other works are centered on users. For instance, in-depth features extracted from tags by means of neural networks are used to represent user profiles in [20]. Preferences of users for specific types of music and user similarities are derived from top tags for tracks and artists in [21], and user expertise is inferred from tags given to music in [7]. There are also proposals in the literature for applying word embedding techniques to tags to improve different aspects of recommender systems [7, 8, 22], but in none of them the tags are used to infer context.

There are hardly any proposals in the literature to infer the context in music listening sessions from social tags. Two papers that do so address the problem of noise produced by redundancy and ambiguity in tags through the creation of clusters. In [4], hierarchical agglomerative clustering of tags is used in a recommendation framework where clusters are selected according to the user's current navigation context. In [5], LDA (Latent Dirichlet Allocation) modelling is used to derive latent topics from the most frequent tags of songs. Then, sequential patterns are induced that allow to predict the topic of the next song to be listened by the user. User contextual information inferred from the sequence of songs is used to provide context-aware recommendations. Another recent work [23] makes use of embedding techniques applied on tags, among other content information, along with contextual information. Convolutional neural networks and other deep learning approaches as attention mechanisms are combined with the word embedding model to derive general and contextual user preferences and make recommendations. In other application domains, some works have been found in which latent contexts are extracted by applying unsupervised deep learning techniques on multiple contextual data that are either obtained from the user explicitly or extracted from sensors [24, 25]. We have not found any work in the literature in which context in user sessions is inferred from tag embedding.

## 3 CONTEXT INFERENCE AND CONTEXT-AWARE RECOMMENDATIONS

This section describes the proposed method for inferring context in user sessions from the music currently being listened to. For this purpose, we make use of social tags associated with the music items. The process for contextual post-filtering applied to provide context-aware recommendations is also described here.

### 3.1 Tag embedding

The processing of the tags consists of treating them with word embedding techniques commonly used in natural language processing. All the tags assigned to the music items by the users of the system are the input to the process. In the same way that words constitute the vocabulary of a given language, social tags form the vocabulary $V$ that will be used to create the model. To obtain the context of the words it is required that these are organized in sentences, but this organization does not exist when we handle tags, thus a way to build sentences must be established. We consider a sentence to include the set of tags that users give to a music item. Given a set of *m* users $U = \{u_1, u_2, \ldots u_m\}$ and a set of *n* items $I = \{i_1, i_2, \ldots i_n\}$, a sentence denoted as $st_j$ is formed by the set of tags $\{t_l\} \subseteq V$ that a subset of users $\{u_i\} \subseteq U$ has given to the item $i_j \in I$. After applying tag embedding to the tags sentences, an output vector $v'_{t_O}$ is obtained for each input tag $t_I$ in the vocabulary $V$, containing the probabilities of each tag to be close to each of the remaining tags.

Word embedding is usually performed by means of two popular techniques: Continuous Bag-of-Words (CBOW) and Skip-gram [26]. For contexts of a given size, the Skip-gram model is trained to predict the probabilities of a word being in the context of the target word, while CBOW predicts the target word from a given context. Both use neural networks to obtain numerical representation of words in documents. The negative log likelihood is used as the loss function to be minimized. The Softmax function is commonly applied as the activation function at the output layer to compute the probability of the output word given the input.

These models can also be used for tag embedding. In this case, the probability $p(t_O|t_I)$ of an output tag $t_O$ given an input tag $t_I$ is computed as follow by using the Softmax function.

$$p(t_O|t_I) = \frac{\exp({v'_{t_O}}^T v_I)}{\sum_{t_I=1}^{V} \exp({v'_{t_O}}^T v_{t_I})} \qquad (1)$$

Our purpose is to use the resulting vectors as indicative of the type of music the user wants to listen to in a particular context. Using dimensionality reduction techniques such as PCA (Principal Component Analysis), tags can be represented in a n-dimensional space such that the closest ones would be those that appear in music items played in similar contexts.

### 3.2 Using tag embedding for contextual postfiltering

The PCA values obtained for the tags are used to select the music items that are recommended to the user in each session. We have opted for a contextual post-filtering approach in which recommendations are generated without considering contextual information by using any collaborative filtering method, either neighborhood-based or matrix factorization, and then those that best fit the user's context at recommendation time are selected. There are two procedures that can be applied for this purpose: Filter the recommendations by discarding those that are not suitable in the target context or readjusting the ranking of the top-N recommendation list to fit the context. Any of the above approaches can be followed to do contextual post-filtering from the PCA values of the tags associated with the items. In order to simplify the process, the reduction to one dimension has been carried out, thus obtaining a single PCA value for each tag.

Since the goal of the proposal is to generate recommendations appropriate to the contextual state in each user session, the first step of the contextual post-filtering is to identify the first song (or songs) played in the session and calculate the average PCA of its tags. This value will be used as a reference to sort the list of recommendations. For this purpose, the re-ranking procedure is followed, ordering the recommended songs according to the distance to the reference PCA, placing first those with the shortest distance.

We start from a set of sessions $S_i = \{s_{i,s}\}$ of the user $u_i \in U$ where each session $s_{i,s}$ is composed by a sequence of items $s_{i,s} = seq\ \{i_{s_{i,s},1}, i_{s_{i,s},2}, \dots, i_{s_{i,s},Q}\}$ such as $i_{s_{i,s},q} \in I$ . Each item is assigned a set of tags with a given PCA value, the average of which is associated with that item: $\overline{PCA}(i_{s_{i,s},q})$. $\overline{PCA}(i_{s_{i,s},1})$ is the reference to be compared with the $\overline{PCA}(i_{s_{i,s},r})$ of the items in the list of top-N recommendations for the session $s_{i,s}$. If $i_{s_{i,s},r}$ is an item of this top-N list, the following difference must be calculated.

$$d\left(i_{s_{i,s},r}\right) = |\overline{PCA}\left(i_{s_{i,s},r}\right) - \overline{PCA}(i_{s_{i,s},1})| \qquad (2)$$

Once the $d\left(i_{s_{i,s},r}\right)$ distances are computed, the top-N list is reordered according to these values from shortest to longest distance.

## 4 EXPERIMENTAL STUDY

### 4.1 Dataset preprocessing

To validate the proposed recommendation approach, an experimental study was conducted using public datasets. One difficulty encountered was the limited availability of datasets with the necessary requirements, that is, information to identify user sessions and calculate ratings, as well as social tags associated with the items.

One of the datasets used in the study was collected by Oscar Celma from the Last.fm streaming platform (https://www.upf.edu/web/mtg/lastfm360k). The required information (user ID, track ID, artist ID and timestamp) was extracted and processed to obtain user sessions from about 420,209 records corresponding to 86,000 songs played by 53 users over two years. In order to identify the songs played in each session and the playing sequence we considered that two consecutive songs played by a user with a time difference of more than 15 minutes belong to different sessions. To get social tags of the items, we used the "Music artists popularity" dataset, obtained from Kaggle (https://www.kaggle.com/), which contains tags of over 1.4 million musical artists collected from MusicBrainz and last.fm. It would have been desirable to have the tags of the songs, but since this was not possible, we used the tags of the artists of these songs to infer the context.

Explicit ratings of the items by users were also missing from the datasets, which led us to derive implicit ratings of the songs for each user from the plays. We used a function of the play frequency percentile proposed by Pacula [27].

### 4.2 Exploratory analysis of the tags

In the proposed method, the user context is inferred from the tags of the items played in the user sessions. In order to determine whether these tags are representative of the context in the music recommendation domain, we have performed a preliminary analysis of them. As mentioned above, tags allow a description of the items in multiple dimensions. In our study, we have found tags referring to genre, year, country, and acoustic characteristics of the music, among other properties, but also referring to different dimensions of the context in which it is listened to. Table 1 shows some examples of tags extracted from the dataset that represent different contexts in which the music to which they are assigned is listened to. As can be seen in the table, the contextual factors or dimensions can be the activity the users are doing, their mood, the place where they are listening to the music, etc. Table 2 shows how tags representative of the type and properties of music, as well as tags representative of different contextual dimensions are used together in the characterization of the items. The latter are shown in bold.

Table 1: Tags representative of different context dimensions

| Contextual dimension | Tags |
|---|---|
| Activity | Dance, listen to while running or riding bike, cooking music, music for fast driving, when studying, good for running in gym, yoga |
| Place | Good gym music, men at work, music fun to hear at work, disco, party, pub rock, lounge at home |
| Mood | Shiny happy music, Melancholic, Sad, Hilarious, Funny, Good mood music, Sentimental |
| Atmosphere | Romantic, quiet, melodic, ambient, dynamic, relax, atmospheric |

Table 2: Examples of tags assigned to items

| Item ID | Tags |
|---|---|
| 4d5d0e4f-8196-428a-957a-665696524792 | Folk, singer-songwriter, australian, Lo-Fi, seen live, indie, acoustic, melbourne, indie folk, asciiecho, **melodic**, songwriter, 2006 explorations, **quiet**, australia, …, **Very chilled**, Boomkat, **slow and sad**, **at work**, waldschrat, awesome and underated. |
| 7ffdf6af-0b1a-4ace-a067-970f0a611d93 | Stand Up, **humour**, Scottish, **funny**, british, folk, …, **disco**, classic rock, 80s, hard rock, **dance**, Stoner Rock, acoustic, album rock, folk-pop, singer, guitar, great, California, bluegrass, male vocalists, 80's, hippie, hippy, romance, **hilarious.** |
| 6bf591c4-b050-478e-9d9f-167394d66c90 | Bossa Nova, …, instrumental; 80s, **ambient, dance**, 90s, summer, r&b, Smooth Jazz, 00s, Jazz Hop, **chillout**, …, **lounge**, funky, **smooth**, euro, cool, check, 50s, 60's, world fusion, tropical, listen to, rio de janeiro, Neo-Soul, 10s, …, **lounge music**, worldmusic, s, lifetime achievement award, **cooking music**, … |

After processing the tags with word embedding techniques, they can be represented in a n-dimensional space by applying PCA to the resulting vectors. The two-dimensional representation for tags with more than 10 occurrences is shown in Figures 1 and 2. The figures show how tags corresponding to very different types of music and contexts are located in distant regions. For example, the area of tags such as “rap rock”, “disco” and “dance-pop” is very separate from the area of tags such as “jazz”, “downtempo” and “chillout”.

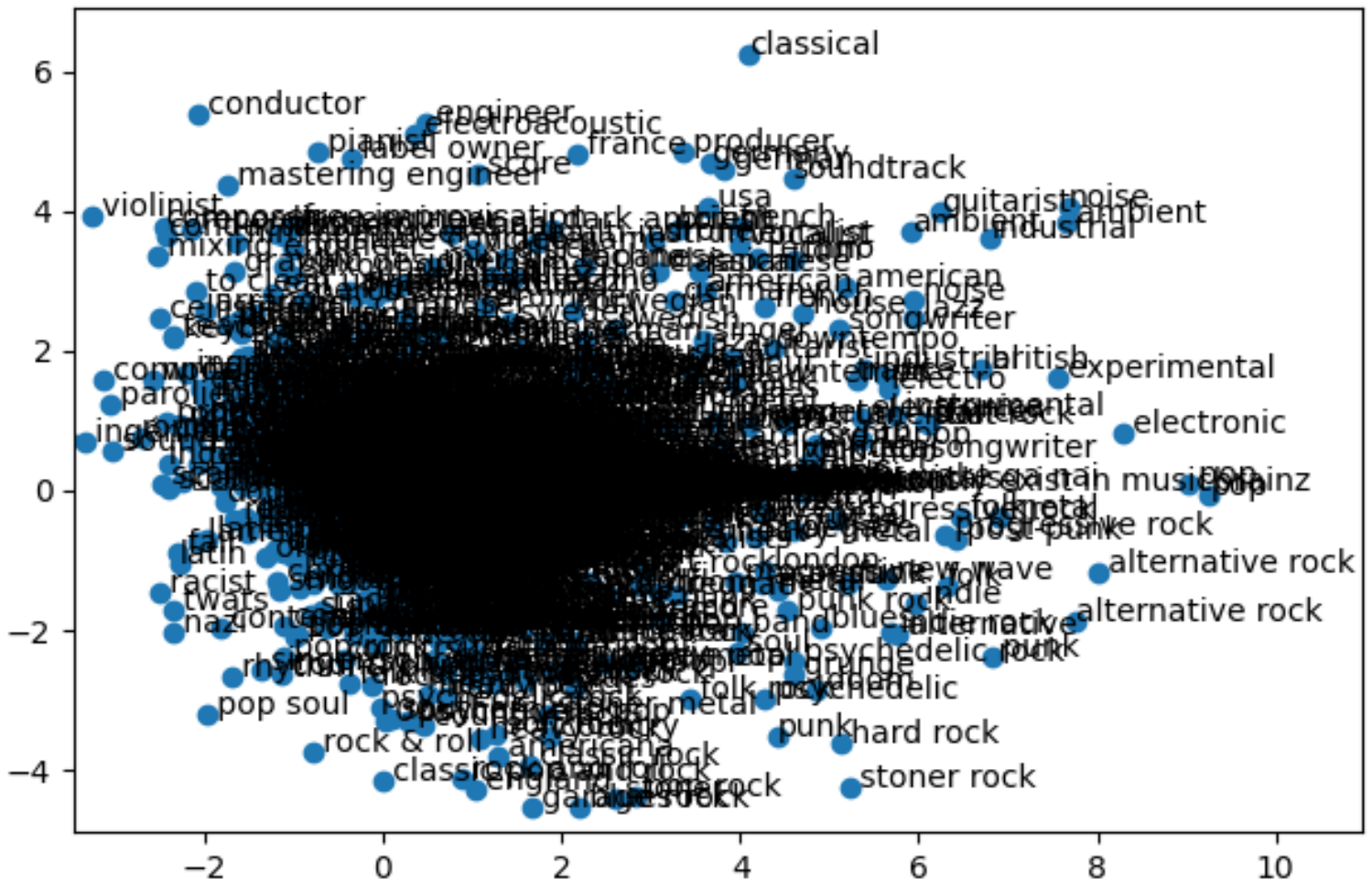

Figure 1: Representation of tag embedding vectors in two-dimensional space.

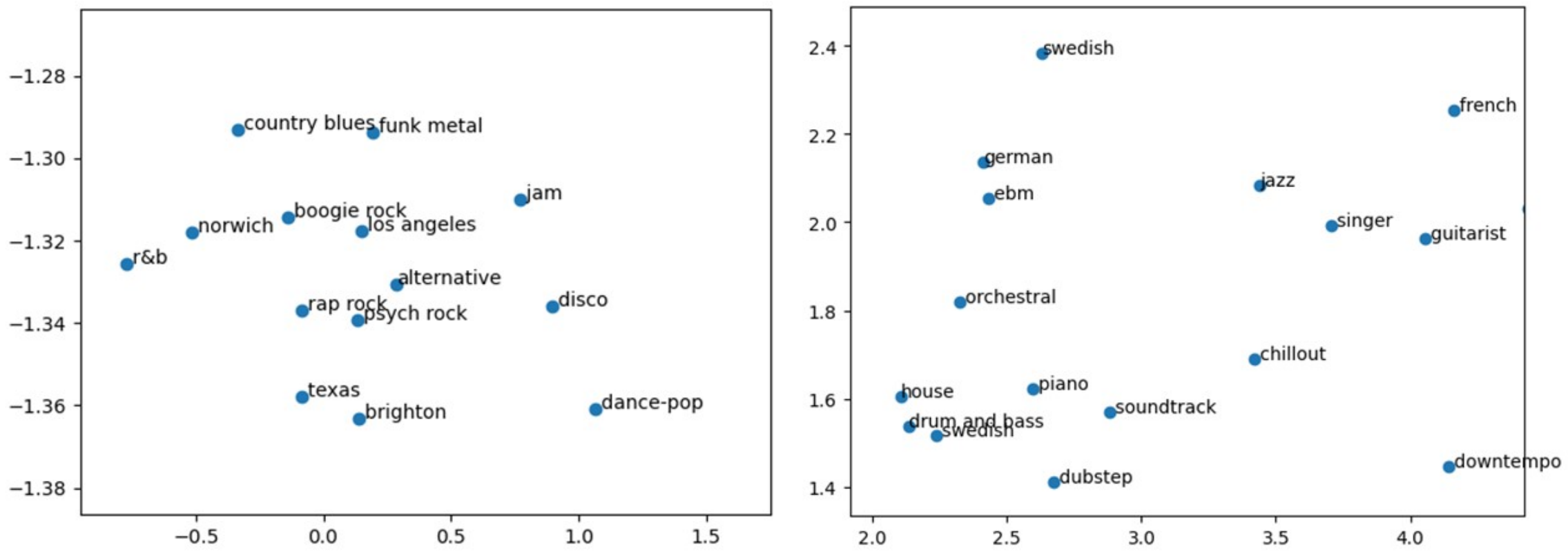

Figure 2: Zoom of two different areas of Figure 1.

### 4.3 Description of experiments and validation

The proposed method was applied to preprocessed data for validation. Previously, word embedding was applied to all the artist tags present in the dataset containing the information of song playing by the users. For this purpose, we used the Gemsim library implementation of Wor2vec. To build the model the min_count parameter was set to 1 aiming at including as many tags as possible, the window size was 5, and epochs=10. The package sklearn.decomposition from the scikit-learn library was used to apply PCA to the output vectors.

The k-NN (k-Nearest Neighbors), SVD (Singular Value Decomposition), SVD++ and NMF (Non-Negative matrix Factorization) methods were applied to obtain the top-N recommendation list. The value of N was set to 40 to obtain the initial list with the best predicted ratings before performing contextual post-filtering, which consisted of reordering it for each session according to the distance of the songs in the list from the starting song in the session (Equation 2). The parameters of the CF methods tested in the study were tuned to obtain the best results.

For validation, the MAP (Mean Average Precision) and NDCG (Normalized Discounted Cumulative Gain) metrics were applied for different values of N. For this purpose, the first N items were selected from the list resulting from post-filtering.

The training and test sets were created by selecting respectively 80% and 20% of the user sessions from the complete dataset. The validation was performed considering each user session individually, comparing the list of songs of each session ordered by rating and the predicted list after post-filtering. That way we can check if the list with contextual post-filtering is more appropriate to the context of each session than the list without post-filtering.

The results of the proposed method were compared with other baseline methods. First, collaborative filtering algorithms (k-NN, SVD, SVD++ and NMF) without contextual post-filtering were applied to test whether contextual post-filtering based on tag embedding improves the recommendations. Secondly, the outputs of these methods were subjected to a post-filtering process based on the similarity of the recommended items with the first played in each session. The objective was to ensure that the improvement of our proposal is really due to the fact that tag-embeddings also enclose the context of the session and not only the music features. To this end, two post-filtering strategies were implemented based on the similarity of the music items. One of them consisted in calculating the similarity of the items based on the ratings assigned to those items by the users. In the other strategy, the similarity of the items was calculated from the TF-IDF of the tags of those items. In both cases, cosine similarity was used.

### 4.4 Results

This section shows the results of the comparative study conducted to validate our proposal against the baselines described above. MAP and NDCG metrics were obtained for top-N lists with N=5, N=10 and N=15 and showed in Figure 3.

The results confirm that the detection of the contextual state in each user session through the embedding of social tags associated with the items leads to a significant increase in the recommendation reliability. The improvements achieved by the proposed postfiltering approach with respect to those obtained with the CF methods without postfiltering range from 38 to 403% in MAP and 557% to 58% in NDCG. The differences with respect to the other two post-filtering methods are also significant although not as high. This can be explained by the fact that the items played in a given context are expected to have a certain similarity. We can also observe that the best results are obtained with the k-NN method both with and without postfiltering. For this method, NDCG achieves values higher than 20% when contextual postfiltering is performed while matrix factorization methods give NDCG values between 4% and 8% approx. With k-NN and post-filtering MAP always exceed 7%, but the rest of the methods after contextual filtering barely reach 4%. The rather low values of the metrics are justified by the fact that the top-N list predicted for each session is compared with the songs played in that session, so the probability of coincidence is much lower than the probability resulting from comparing the lists for a given user regardless of the sessions.

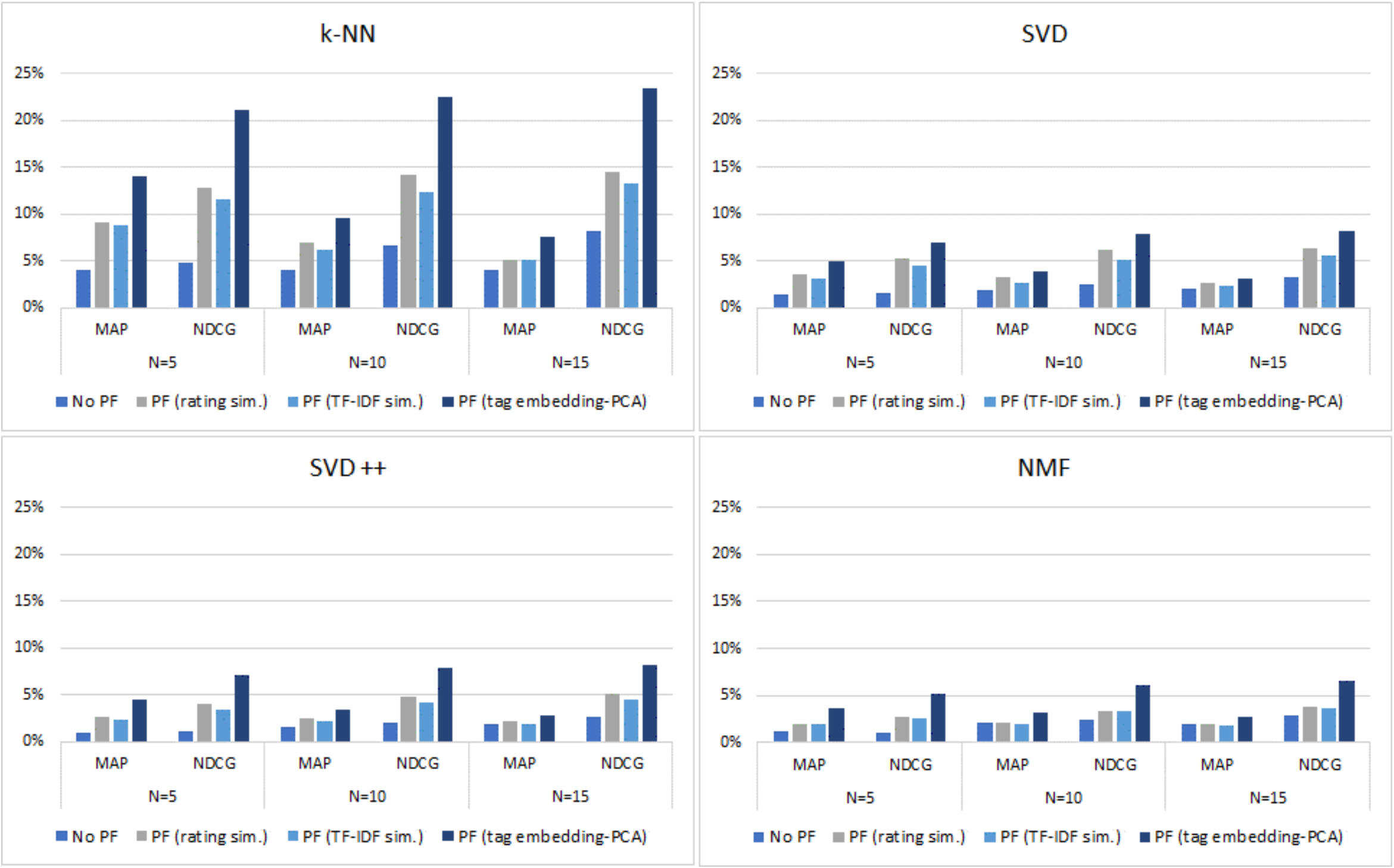

Figure 3: Results of the k-NN, SVD, SVD++ and NMF methods applied in four different ways: without postfiltering (“No PF”), with postfiltering using rating-based item similarity (“PF (rating sim.)”), with postfiltering using TF-IDF-based item similarity (“PF (TF-IDF sim.)”) and by means of postfiltering based on tag embedding (“PF (tag embedding-PCA)”)

## 5 CONCLUSIONS

User context inference is a topic of great interest in music recommender systems due to the influence of contextual factors on user preferences as well as the difficulty of eliciting them both explicitly and implicitly. In this work we have made use of the social tags that users assign to music items to determine the contextual state of the user in each listening session. The process involves the application of word embedding techniques to the tags to obtain the vectors with the probability of each one of them to be close to each of the others. The PCA dimensionality reduction technique is applied to the output vectors. Thus, the less distance there is between items in terms of the PCA of their tags, the more similar will be the context in which they are played. That distance between the first item(s) of the session and the rest of the items in the top-N list produced by any collaborative filtering method is used to do contextual post-filtering and recommend those items that best fit the user context in that session. The results of the experimental study conducted show a significant improvement in the reliability of the recommendations when the proposed method is applied in relation to those obtained without taking into account the context or using other post-filtering strategies.

## ACKNOWLEDGMENTS

This research has been supported by the Department of Education of the “Junta de Castilla y León”, Spain, (ORDEN EDU/667/2019). Grant no: SA064G19.